# Kirigami-based Elastic Metamaterials with Anisotropic Mass Density for Subwavelength Flexural Wave Control

R. Zhu[1,2*], H. Yasuda[1], G.L. Huang[3] and J.K. Yang[1*]

[1] *Department of Aeronautics and Astronautics, University of Washington, Seattle, WA 98195, USA*
[2] *School of Aerospace Engineering, Beijing Institute of Technology, Beijing 100081, People's Republic of China*
[3] *Department of Mechanical & Aerospace Engineering, University of Missouri, Columbia, MO 65211, USA*

## ABSTRACT

A novel design of an elastic metamaterial with anisotropic mass density is proposed to manipulate flexural waves at a subwavelength scale. The three-dimensional metamaterial is inspired by kirigami, which can be easily manufactured by cutting and folding a thin metallic plate. By attaching the resonant kirigami structures periodically on the top of a host plate, a metamaterial plate can be constructed without any perforation that degrades the strength of the pristine plate. An analytical model is developed to understand the working mechanism of the proposed elastic metamaterial and the dispersion curves are calculated by using an extended plane wave expansion method. As a result, we verify an anisotropic effective mass density stemming from the coupling between the local resonance of the kirigami cells and the global flexural wave propagations in the host plate. Finally, numerical simulations on the directional flexural wave propagation in a two-dimensional array of kirigami metamaterial as well as super-resolution imaging through an elastic hyperlens are conducted to demonstrate the subwavelength-scale flexural wave control abilities. The proposed kirigami-based metamaterial has the advantages of no-perforation design and subwavelength flexural wave manipulation capability, which can be highly useful for engineering applications including non-destructive evaluations and structural health monitoring.

---
*Email: rzhu83ac@gmail.com
*Email: jkyang@aa.washington.edu

Elastic metamaterials (EMMs) with delicately engineered architectures possess abnormal effective material properties and therefore, can control elastic wave propagation in a subwavelength scale [1-8]. Anisotropic mass density (AMD), achieved by the EMMs, can easily convert stress waves emitting from an omnidirectional source into highly directional wave propagation, which is of major interest in structural health monitoring (SHM). Since the wave energy can be sent in a preferential direction, an improved damage sensitivity can be expected due to the increased interaction between the damage and the interrogation signal. Most recently, an elastic hyperlens – built by the EMM with AMD or anisotropic stiffness – has grasped significant attention with a breakthrough in overcoming the diffraction limit and realizing super-resolution damage imaging even in the far field [9-11]. Although the EMM with AMD opens a new avenue to dramatically improving both damage sensitivity and imaging resolution, which are one of the most challenging problems in a SHM system, there still exist difficulties lying in their practical design and applications.

Currently, the EMM designs to achieve AMD require either embedded complex resonant inclusions with a combination of soft and rigid material components or perforated/engraved single-phase structures. Zhu *et al.* designed and experimentally validated a EMM plate consisting of embedded anisotropic resonant inclusions made of lead cylinders coated with elliptical epoxy layers where drilled holes were introduced to further increase the anisotropicity in its effective mass density tensor[12]. A single-phase version of the EMM plate was later designed by them, which only requires two laser-cut slots in each unit cell and AMD was numerically demonstrated for directional longitudinal wave propagation and elastic hyperlens[11]. Li *et al.* designed and experimentally tested a metamaterial hyperlens with strongly AMD which was realized in a brass plate[13]. Torrent and Sanchaz-Dehesa also achieved AMD with fluid in a planar wave guide made by drilling circular cavities in an aluminum plate[14]. Dong *et al.* applied topology optimization on hyperbolic EMM made of perforated single-phase metal microstructures to achieve broadband AMD and super-resolution imaging of longitudinal elastic waves[15]. From a practical point of view, these subtractive metamaterial designs inevitably decrease the strength of the pristine plate and therefore, are not suitable for real-world SHM applications since they introduce damages in the first place. Moreover, most studies on the EMMs with AMD focus primarily on the propagation of bulk waves in the supra-wavelength scales. The design of the anisotropic EMMs to control subwavelength flexural wave propagation has not been fully explored, despite their usefulness in a number of elastic wave devices[16-19] and guided wave based damage detection systems[20-22].

Attachable structural units, additively built on the surfaces of a plate-like EMM, offer a damage-free and therefore, more practical solution to control flexural waves which, comparing with in-plane waves, have higher sensitivity and better damage imaging resolution. Xiao *et al.* analytically studied the flexural wave propagation in a Kirchhoff plate attached with a 2D periodic array of spring-mass resonators[23]. Hsu *et al.* numerically investigated a locally resonant EMM plate composed of periodically attached stepped resonant stubs[24]. Yan *et al.* proposed a EMM flat lens by bonding a 2D planar array of lead-rubber composite cylinders on the aluminum thin plate for the $A_0$-mode Lamb wave focusing[19]. Colombi *et al.* designed an EMM plate by attaching vertical narrow cylinders to a thin plate and realized energy trapping[25]. Those highly symmetric surface-bonded structures naturally provide isotropic coupling between the resonator and the bottom plate and therefore, cannot create desired anisotropic coupling behavior in a subwavelength scale. In order to achieve the AMD for flexural wave propagation in an EMM plate, distinctive couplings between the motion of the built-on-top resonators and the out-of-plane motion in the bottom plate need to be generated along different wave propagation directions. Also, a single-phase metallic resonant structure is preferred since soft materials may introduce damping which weakens the coupling between the top resonators and the bottom plate.

Kirigami ('kiri' means cut), similar to origami, is a paper folding technique that involves both 'cut' and 'fold' processes and becomes an emerging research frontier due to its ability to transform 2D flat paper/metal sheets into three dimensional (3D) structures with targeted functions. Such 3D structures with tunable fundamental properties can further lead to potential applications in design of stretchable and conformable electronics[26], morphing architectures[27], and reprogrammable mechanical and material systems[28]. These origami/kirigami-inspired engineering designs have demonstrated that they not only create complex 3D morphology but also give birth to the desired static as well as dynamic characteristics. Naturally, one can expect improved and even new dynamic behavior and wave functions from the origami/kirigami-inspired metamaterials. Overvelde *et al.* exploited an origami-inspired programmable mechanical metamaterial, whose shape, volume, and stiffness can be actively controlled[29]. Babaee *et al.* designed and experimentally tested a new class of reconfigurable acoustic waveguides based on 3D origami-inspired metamaterials[30]. Harne and Lynd investigated the topological reconfigurations of a Miura-ori-based acoustic array to achieve acoustic wave energy focusing[31]. Yasuda *et al.* investigated the nonlinear wave dynamics of origami-based metamaterials composed of Tachi-Miura polyhedron unit cells which can be tuned by modifying their geometrical configurations or initial folded conditions[32]. Extending the advantages of the kirigami/origami technique to the subwavelength structure in EMM can bring new design possibility as well as advantageous 3D fabrication means. Particularly, applying the kirigami/origami technique to the realizations of an attachable EMM with AMD and flexural wave manipulation without compromising the pristine plate strength is of practical importance.

In this study, an attachable EMM microstructure has been designed based on the kirigami technique to achieve AMD for subwavelength flexural wave propagation in a thin plate. The AMD of the kirigami EMM was obtained numerically by using the effective medium method, and the frequency band structure of the anisotropic flexural wave propagation was calculated from the developed mass-spring-plate model. The agreement in the AMD frequency region can be found between the results of the numerical method and the analytical model. Finally, the directional flexural wave propagation and the attachable super-resolution elastic hyperlens were demonstrated numerically. The proposed kirigami-based EMM with AMD has the advantages of easy fabrication, damage-free design, and subwavelength flexural wave manipulation, which are highly desirable features for the engineering applications, such as non-destructive evaluations and SHM.

## Results

**Design of kirigami-based elastic metamaterial.** Figure 1 shows the schematic of the proposed kirigami cell design. We first cut a 0.4 mm-thick aluminum (Al) thin plate (Fig. 1a) and then fold it along the dashed lines (Fig. 1b) to form a 3D structure (Fig. 1c). The 3D Al structure is attached onto a 1 mm-thick Al substrate with a heavy mass block (18 g) glued on its top (Fig. 1d). The lattice constant of the kirigami-plate unit cell is $a$. The geometrical parameters and the material properties of the kirigami cell is listed in Table 1. It should be noted the shapes and widths of the kirigami legs along the $x$- and $y$-directions are made different intentionall. As a result, different resonant frequencies can be expected for the flexural wave propagations along the two principle directions of the EMM. The legs with a larger width and a U-shaped cross section possess larger bending stiffness and therefore, introduce high resonant frequency along the $y$-direction. On the contrary, the other two legs with a smaller width and a flat-rectangle cross section have smaller bending stiffness and therefore, reduce the $x$-directional resonant frequency. It should be mentioned that the size of the mass attached on top of the kirigami structure can be designed to lower both $x$- and $y$-directional resonant frequencies into a desired frequency regime in order to ensure that the EMM unit cell vibrates in a subwavelength scale.

The effective mass density of the proposed kirigami EMM can be calculated by using a computational-based effective medium method[33]. By applying out-of-plane displacement excitations $W_x$ and $W_y$ along the $x$ and $y$ directions, the resultant forces of the metamaterial unit cell along the two directions can be numerically obtained as $F_{Z,x}$ and $F_{Z,y}$, respectively. Thus, the AMD tensor composed of the x- and y-directional effective mass densities, $\rho_{eff,x}$ and $\rho_{eff,y}$, can be readily determined from the following relation[11,33]

$$\begin{bmatrix} F_{Z,x} \\ F_{Z,y} \end{bmatrix} = -\omega^2 V \begin{bmatrix} \rho_{eff,x} & 0 \\ 0 & \rho_{eff,y} \end{bmatrix} \begin{bmatrix} W_x \\ W_y \end{bmatrix} \qquad (1)$$

where $\omega$ represents the excitation angular frequency, and $V$ denotes the volume of the unit cell.

Figure 2 shows the dynamic $\rho_{eff,x}$ and $\rho_{eff,y}$ of the attached kirigami structure as a function of $\omega$, given the parameters listed in Table 1. Here, $\rho_{eff,x}$ and $\rho_{eff,y}$ values are normalized with the static effective mass density of the EMM unit cell (i.e., $\rho_{eff,x}$ and $\rho_{eff,y}$ at $\omega = 0$). In Fig. 2, it is evident that this kirigami structure exhibits the anisotropy in its effective mass densities, though the general trends of $\rho_{eff,x}$ and $\rho_{eff,y}$ are similar. Again, the discrepancy of these two curves stems from the distinctive leg stiffness of the kirigami structure. Also, it can be found that $\rho_{eff,x}$ and $\rho_{eff,y}$ have different signs in the frequency range from $f = 5630$ Hz to $f = 7150$ Hz, which is highlighted in gray. This is an interesting feature which will eventually enable us to achieve the directional propagation of flexural waves. Lastly, in this shaded frequency range, the wavelength of the A0-mode flexural wave propagating in the thin substrate structure is approximatley $\lambda = 41$mm at 6 kHz. This wavelength corresponds to almost four times of the characteristic size of the kirigami cell represented by the lattice constant ($a = 10$ mm in this study). This manifests the subwavelength nature of the designed kirigami cell. We will verify these unique features of the kirigami-based EMM in the following part of the manuscript.

A mass-spring-plate model was also developed to analyze the band structures of the kirigami metamaterial, as shown in Fig. 3a. Consider an infinitely thin plate lying in the $x$-$y$ plane with periodically attached resonators consisting of two pairs of springs $k_x$ and $k_y$ and a top mass $m_R$, as shown in Fig. 3b. Here, $a$ is the lattice constant of the unit cell, and $a_x$ and $a_y$ are the distances between the two points where the springs with coefficients $k_x$ and $k_y$ are attached along the $x$-and $y$-directions, respectively. Inside the unit cell, the four attachment points can be denoted in vector forms with the origin being at the center of the unit cell, as shown in Fig. 3c:

$$\vec{H}_1 = 0.5a_x\hat{x} + 0\hat{y} \,; \vec{H}_2 = -0.5a_x\hat{x} + 0\hat{y}$$
$$\vec{V}_1 = 0\hat{x} + 0.5a_y\hat{y} \,; \vec{V}_2 = 0\hat{x} - 0.5a_y\hat{y} \,. \tag{2}$$

The detailed derivations of the governing equations and the extended plane wave expansion method for band structure calculation are in the Methods section.

The band structures along both $x$-and $y$-directions are calculated with the two spring constants being determined by conducting finite element-based static analysis on the kirigami legs. As a result, $k_x = 9.5$ MPa and $k_y = 21$ MPa are retrieved from the force-displacement curves, as shown in Figs. 4a and b. The band structures obtained from the mass-spring-plate model are plotted in Fig. 4c alongside with the effective mass densities. It can be found that the frequency difference between the $x$-directional and the $y$-directional band gap (light gray shaded area) overlaps with the frequency range where $\rho_{eff,x}$ and $\rho_{eff,y}$ have different signs (light yellow shaded area).

Note that it is not only the difference between the two spring constants $k_x$ and $k_y$, but also the difference between the $a_x$ and $a_y$ that plays an important role in creating the AMD in the metamaterial unit cell. In our proposed kirigami structure, the difference between $k_x$ and $k_y$ indicates different geometric shapes and thicknesses of the kirigami legs, while the difference between $a_x$ and $a_y$ represents variations in bending angles and leg-attached positions on the bottom plate. Figure 5 shows the effect of different parameter ratios ($k_x/k_y$ and $a_x/a_y$) on the band structure of the anisotropic kirigami metamaterial. Here, $\Delta\omega^*$ is the width of the strong-AMD frequency band (i.e., highlighted frequency band in Fig. 4c) where $\rho_{eff,x}$ and $\rho_{eff,y}$ have opposite signs which is the critical condition to realize hyperlens. It can be found that $\Delta\omega^*$ increases dramatically when either $k_x/k_y$ or $a_x/a_y$ starts to increase from a unity. However, when the difference between the spring coefficients or their attachment locations is large enough, $\Delta\omega^*$ increases very slowly. It is also notice that increasing $a_x/a_y$ is more efficient in creating large strong-AMD region with $\Delta\omega^*$ being up to 0.22 while increasing $k_x/k_y$ brings a slower change in $\Delta\omega^*$ and a relatively smaller strong-AMD region with $\Delta\omega^*$ being up to 0.14.

**Directional flexural wave propagation.** Now we demonstrate numerically that the proposed kirigami-based EMM can propagate flexural waves directionally by leveraging the anisotropic effective mass density explained in the previous section. Figures 6 a and b show full-scale simulation results of elastic waves propagating in a regular thin plate and a kirigami-decorated plate, respectively (software: COMSOL). In both cases, a pair of piezoelectric disks (PZT-5H) are surface-bonded at the center of the plate, with one on the top and the other on the bottom of the plate. Same amplitude but opposite signs of the voltage excitations are applied to the PZT pair for the asymmetric deformation, which excites purely flexural wave (A0-mode Lamb wave) in the plate[35]. Perfectly matched layers (PMLs) are built around the plate region to eliminate any unwanted boundary reflections. Without the kirigami array attached, the flexural wave ($f$ = 6.15 kHz) propagates isotopically in the thin Al plate, as shown in Fig. 6 a. With the kirigami array attached around the PZT disk on the top surface of the plate, the flexural wave propagates only along the $y$-direction in the bottom plate, as shown in Fig. 6 b. This is because $\rho_{eff,y} > 0$, while $\rho_{eff,x} < 0$ in the formed metamaterial plate as shown in Fig. 2. That is, the positive effective mass density supports the propagation of waves, while the negative one forms evanescent waves that decay exponentially in space. We again note in passing that the characteristic length of the kirigami cell (10 mm) is only ¼ of the flexural wavelength in the bottom plate ($\lambda = 41$ mm). Such capability of the subwavelength wave management can be further exploited for applications such as the super-resolution elastic hyperlens as demonstrated below. The left and right parts of Fig. 6c show the flexural wave fields at $f = $ 2 kHz and 8 kHz, respectively. More flexural wave control abilities of the attached kirigami array can be found with near-isotropic propagation at 2 kHz and all-direction wave block at 8 kHz, which are results of all-positive densities $\rho_{eff,x} \approx \rho_{eff,y} > 0$ and all-negative densities $\rho_{eff,y} < \rho_{eff,x} < 0$, respectively.

**Demonstration of flexural wave hyperlens.** Elastic hyperlens, designed by the EMM with the mass density being positive in one direction while negative in the orthogonal direction, can not only permit the transmission of the evanescent waves carrying the subwavelength information to the near field, but also magnify it, thereby converting it to propagating waves in the far field[36,37]. Although previous acoustic and elastic/acoustic hyperlenses have been demonstrated for the subwavelength pressure waves[13] and in-plane longitudinal elastic waves[9-11], respectively, the hyperlens for flexural wave has not been fully explored due to the lack of proper subwavelength architectures that can lead to the required AMD. Our proposed kirigami structure can not only generate the AMD for flexural waves but also provide a surface-attached hyperlens solution that can be the first step towards the development of a scanning super-resolution damage imaging system.

Figure 7 shows the design of the attachable elastic hyperlens with the proposed kirigami structures. The 180° sector-shaped hyperlens consists of 115 kirigami units, which are arranged in seven layers (the numbers of the units in each layer are marked in Fig. 7a). The inner and outer radii of the hyperlens are 18 mm and 106 mm, respectively, which indicates a magnification ratio of six for the enlarged subwavelength feature. The distance between neighboring unit cells along the circumferential direction is the same as that along the radial direction. To demonstrate its super-resolution imaging ability for flexural waves, two pairs of 2 mm-radius piezo disks are bonded to the top and bottom surfaces of the plate with their voltage excitations tuned for dominated flexural wave generation. The distance between the two disks on the top surface of the plate is 13 mm, which is less than 1/3 λ at 6.15 kHz. Thus, this subwavelength separation of the excitation sources cannot be detected in the regular thin Al plate without attached kirigami hyperlens (Fig. 7b).

Figure 7c shows the simulation result of the kirigami-decorated plate, given the same excitation condition described above. Here, the distance between the two sources and the first layer of the hyperlens is 1.8 mm, which is kept small enough for the lens to pick up the evanescent wave that carries the subwavelength feature information. Thanks to AMD, the evanescent wave reaching the inner array of the kirigami cells is converted to the propagating wave that is transmitted the outer region along the radial direction, as shown in Fig. 7c. Therefore, the subwavelength information of the two wave sources can be channeled out to the outer boundary of the hyperlens without significant attenuation. Meanwhile, the distance between the two wave sources is enlarged six times as defined by the magnification ratio, which enables the far field super-resolution imaging since the enlarged feature is now larger than the diffraction limit. Note that this super-resolution capability is achieved only in the frequency range of the AMD. For example, Fig. 7d shows the simulation results at $f = 8$ kHz where both $\rho_{eff,x}$ and $\rho_{eff,y}$ have negative values.

Therefore, both propagating and evanescent flexural waves generated by the two wave sources are totally blocked by the attached kirigami hyperlens.

**Discussion**

In this paper, an attachable elastic mechanical metamaterial (EMM) design has been proposed based on kirigami to control subwavelength flexural waves propagating in a thin substrate structure. By strategic design of the kirigami cells, this EMM structure can exhibit the anisotropic mass density (AMD). We calculated the AMD of the kirigami EMM by the numerical-based effective medium method. We also obtained the frequency band structure of the anisotropic flexural wave propagation by using a lumped mass-spring-plate model. Good agreement in the strong-AMD frequency region can be found between the results of the numerical method and the analytical model. Based on this, the directional flexural wave propagation and the concept of the attachable super-resolution elastic hyperlens were demonstrated numerically. The proposed kirigami-based EMM with the AMD has the advantages of no-perforation design and subwavelength flexural wave manipulation capability, which can be highly useful features for engineering applications in the fields of non-destructive evaluations and SHM.

**Methods**

**Extended plane wave expansion method for mass-spring-plate model.** By using the Kirchhoff plate theory and the Dirac delta function to couple the flexural wave motion of the bottom plate and the vertical vibrations of the attached resonators, the governing equation of the metamaterial plate can be written as

$$\begin{cases} D\nabla^4 W(\vec{r}) - \omega^2 \rho h W(\vec{r}) = \Sigma_R \left( f_{H_1}\delta_{H_1} + f_{H_2}\delta_{H_2} + f_{V_1}\delta_{V_1} + f_{V_2}\delta_{V_2} \right) \\ -\omega^2 m_R U_z(\vec{R}) = -\left( f_{H_1} + f_{H_2} + f_{V_1} + f_{V_2} \right) \end{cases} \quad (3)$$

with

$$f_{H_1} = -k_x[W(\vec{R}+\vec{H}_1) - U_z(\vec{R})], \; f_{H_2} = -k_x[W(\vec{R}+\vec{H}_2) - U_z(\vec{R})]$$
$$f_{V_1} = -k_y[W(\vec{R}+\vec{V}_1) - U_z(\vec{R})], \; f_{V_2} = -k_y[W(\vec{R}+\vec{V}_2) - U_z(\vec{R})] \quad (4)$$

and

$$\delta_{H_1} = \delta[\vec{r} - (\vec{R}+\vec{H}_1)], \; \delta_{H_2} = \delta[\vec{r} - (\vec{R}+\vec{H}_2)]$$
$$\delta_{V_1} = \delta[\vec{r} - (\vec{R}+\vec{V}_1)], \; \delta_{V_2} = \delta[\vec{r} - (\vec{R}+\vec{V}_2)] \quad (5)$$

Where $\rho$ and $h$ are the mass density and thickness of the bottom plate; $\vec{r}$ is the position vector in the bottom plate; $\vec{R} = ma\hat{x} + na\hat{y}$ is introduced to describe the position of the unit cell in the $n^{th}$ row and $m^{th}$ column of the metamaterial plate; $W$ and $U_z$ present the out-of-plane displacement of the bottom plate and the attached resonator, respectively. The bottom plate has the bending stiffness $D = Eh^3/12(1-v^2)$, where $E$ and $v$ represent the Young's modulus and the Poisson's ratio of the plate.

To form an eigenvalue problem from governing equation Eq. (3) and obtain the band structure of the metamaterial plate, the plane wave expansion method [34] is extended to deal with the continuous plate system with discretely positioned spring-mass resonant elements. The out-of-plane displacement $W$ of the bottom plate can be assumed as

$$W(\vec{r}) = \sum_G W_0(\vec{G}) e^{-i(\vec{k}+\vec{G})\cdot\vec{r}} \quad (6)$$

Where $W_0$ is the corresponding Fourier coefficients of the displacement $W$, $\vec{k}$ is the flexural wave vector and $\vec{G}$ is the reciprocal-lattice vector as:

$$\vec{G} = \tilde{m}\hat{b}_x + \tilde{n}\hat{b}_y \quad (7)$$

with $\hat{b}_x = (2\pi/a_x, 0)$ and $\hat{b}_y = (0, 2\pi/a_y)$ being the basis vectors of the reciprocal lattice, $\tilde{m}$ and $\tilde{n}$ being an integer. Here, $a_x$ and $a_y$ are the distances between the spring attachment points along the $x$-and $y$-directions, respectively (Fig. 3c). By applying the periodic condition:

$$W(\vec{R}) = W(0)e^{-i\vec{k}\cdot\vec{R}}, U_z(\vec{R}) = U_z(0)e^{-i\vec{k}\cdot\vec{R}}$$

$$W(\vec{R} + \vec{H}_1) = W(\vec{H}_1)e^{-i\vec{k}\cdot\vec{R}}, W(\vec{R} + \vec{H}_2) = W(\vec{H}_2)e^{-i\vec{k}\cdot\vec{R}} \quad (8)$$

$$W(\vec{R} + \vec{V}_1) = W(\vec{V}_1)e^{-i\vec{k}\cdot\vec{R}}, W(\vec{R} + \vec{V}_2) = W(\vec{V}_2)e^{-i\vec{k}\cdot\vec{R}}$$

The governing equation Eq. (3) can be rewritten as

$$DS\left[(k_x + G_x)^2 + (k_y + G_y)^2\right]^2 \sum_{\vec{G}_0} \widetilde{W}(\vec{G}_0)e^{-i(\vec{k}+\vec{G}_0)\cdot\vec{r}} - \omega^2 \rho h S \sum_{\vec{G}_0} \widetilde{W}(\vec{G}_0)e^{-i(\vec{k}+\vec{G}_0)\cdot\vec{r}}$$

$$= -2k_x \left\{\sum_{\vec{G}_0} \widetilde{W}(\vec{G}_0)\cos[0.5a_x(G_{0x} - G_x)] - U_z(0)\cos[0.5a_x(k_x + G_x)]\right\}\left[\sum_{\vec{G}_0} e^{-i(\vec{k}+\vec{G}_0)\cdot\vec{r}}\right]$$

$$- 2k_y \left\{\sum_{\vec{G}_0} \widetilde{W}(\vec{G}_0)\cos[0.5a_y(G_{0y} - G_y)] - U_z(0)\cos[0.5a_y(k_y + G_y)]\right\}\left[\sum_{\vec{G}_0} e^{-i(\vec{k}+\vec{G}_0)\cdot\vec{r}}\right] \quad (9)$$

$$-\omega^2 m_R U_z(0) = 2k_x \left\{\sum_{\vec{G}_0} \widetilde{W}(\vec{G}_0)\cos[0.5a_x(k_x - G_{0x})] - U_z(0)\right\}$$

$$+ 2k_y \left\{\sum_{\vec{G}_0} \widetilde{W}(\vec{G}_0)\cos[0.5a_y(k_y - G_{0y})] - U_z(0)\right\}$$

It should be mentioned that the relation $\sum_{\vec{R}} \delta(\vec{r} - \vec{R}) = \frac{1}{S}\sum_{\vec{G}} e^{-i\vec{G}\cdot\vec{r}}$ with $S$ being the area of the unit cell is used in the derivation of Eq. (9). By removing $\sum_{\vec{G}_0} e^{-i(\vec{k}+\vec{G}_0)\cdot\vec{r}}$ from both sides of the equation, Eq. (9) can eventually be represented in a matrix form as:

$$\begin{bmatrix} \rho h S[I] & \{0\} \\ \{0\}^T & m_R \end{bmatrix}^{-1} \begin{bmatrix} DS[K] + 2k_x[CGGX] + 2k_y[CGGY] & -2k_x[CGGX] - 2k_y[CKGY] \\ -2k_x\{CKGX\}^T - 2k_y\{CKGY\}^T & 2(k_x + k_y) \end{bmatrix} \begin{Bmatrix} \{\widetilde{W}\} \\ U_z \end{Bmatrix}$$

$$= \omega^2 \begin{Bmatrix} \{\widetilde{W}\} \\ U_z \end{Bmatrix} \quad (10)$$

where

$$[K]$$
$$= \begin{bmatrix} \left[\left(k_x + G_{1_x}\right)^2 + \left(k_y + G_{1_y}\right)^2\right]^2 & \cdots & 0 \\ \vdots & \ddots & \vdots \\ 0 & \cdots & \left[\left(k_x + G_{(2\widetilde{m}+1)^2 x}\right)^2 + \left(k_y + G_{(2\widetilde{m}+1)^2 y}\right)^2\right]^2 \end{bmatrix}_{(2\widetilde{m}+1)^2 \times (2\widetilde{m}+1)^2}$$

$$[CGGX] = \begin{bmatrix} \cos\left[\frac{a_H}{2}(G_{1x} - G_{1x})\right] & \cdots & \cos\left[\frac{a_H}{2}(G_{(2\widetilde{m}+1)^2 x} - G_{1x})\right] \\ \vdots & \ddots & \vdots \\ \cos\left[\frac{a_H}{2}(G_{1x} - G_{(2\widetilde{m}+1)^2 x})\right] & \cdots & \cos\left[\frac{a_H}{2}(G_{(2\widetilde{m}+1)^2 x} - G_{(2\widetilde{m}+1)^2 x})\right] \end{bmatrix}_{(2\widetilde{m}+1)^2 \times (2\widetilde{m}+1)^2} \quad (11)$$

$$\{CKGX\} = \begin{Bmatrix} \cos\left[\frac{a_H}{2}(k_x + G_{1x})\right] \\ \vdots \\ \cos\left[\frac{a_H}{2}(k_x + G_{(2\widetilde{m}+1)^2 x})\right] \end{Bmatrix}_{(2\widetilde{m}+1)^2 \times 1}, [I] = \begin{bmatrix} 1 & \cdots & 0 \\ \vdots & \ddots & \vdots \\ 0 & \cdots & 1 \end{bmatrix}_{(2\widetilde{m}+1)^2 \times (2\widetilde{m}+1)^2},$$

$$\{\widetilde{W}\} = \begin{Bmatrix} \widetilde{W}_1 \\ \vdots \\ \widetilde{W}_{(2\widetilde{m}+1)^2} \end{Bmatrix}_{(2\widetilde{m}+1)^2 \times 1}$$

where we choose $m, n = (-\widetilde{m}, \cdots, \widetilde{m})$ meaning that $(2\widetilde{m} + 1)^2$ plane waves are used in the calculation. Eq. (10) is a generalized eigenvalue problem for $\omega^2$. By solving this equation for each flexural wave vector $\vec{k}$, $(2\widetilde{m} + 1)^2 + 1$ eigenvalues in terms of $\omega^2$ can be obtained, which can lead to the flexural wave band structures of the kirigami metamaterial plate.

## Acknowledgments

R. Z., H. Y., and J.K.Y. are grateful for the financial support from U.S. Office of Naval Research (N00014140388) and NSF(CAREER-1553202). R. Z. acknowledges the support from the Thousand Young Talents Program of China. G. L. acknowledges the support from U.S. Air Force Office of Scientific Research (AF9550-15-1-0016).

## Author Contributions

R. Z. and J.K.Y. conceived the idea of the project. R.Z. and H.Y. design the kirigami cell; R.Z. conducted the analytical and numerical simulations. R.Z., J.K.Y. and G.L.H. wrote the manuscript. J.K.Y. supervised the project.

## Competing financial interests

The authors declare no competing financial interests.

## Figure legends

**Figure 1. The schematic of the kirigami structure design following the cut-fold principle.** (**a**) Cut on a thin Al plate. (**b**) Fold the plate along the dashed lines. (**c**) The 3D folded kirigami cell. (**d**) Attach the kirigami cell on a bottom plate with glued mass block.

**Figure 2. The dynamic effective mass densities of the kirigami metamaterial cell**.

**Figure 3. Mass-spring-plate model.** (**a**) Schematic diagram of the mass-spring-plate model. (**b**) A single unit cell. (**c**) The locations of the four spring attachment points in the unit cell.

**Figure 4. Effective spring constants and band structures of the mass-spring-plate model.** (**a**) $k_x$ retrieved from the displacement-force curves based the finite element simulations. (**b**) Retrieved $k_y$. (**c**) Comparison between band structures in the *x*- and *y*-directions and their relation to the effective mass density diagram. The insets show the von Mises stress profiles.

**Figure 5.** The effect of the parameter ratios ($k_x/k_y$ or $a_x/a_y$) on $\Delta\omega^*$

**Figure 6. Numerical simulations of the directional wave propagation in kirigami EMM plate. (a)** Isotropic flexural wave propagation in a pure Al thin plate. (**b**) Directional flexural wave propagation in an Al thin plate with the attached kirigami EMM array. The images in the left part of (a) and (b) illustrate geometric configurations, while the right parts show the out-of-plane displacements of the substrate represented in colors. (**c**) Flexural wave propagations in EMM plate at 2kHz and 8kHz.

**Figure 7. Numerical simulations of the elastic hyperlens for flexural waves. (a)** Design of kirigami elastic hyperlens for flexural waves. (**b**) Flexural wave field without attached

hyperlens at 6.15kHz. **(c)** Flexural wave field with attached hyperlens at 6.15kHz. **(d)** Flexural wave field with attached hyperlens at 8kHz.

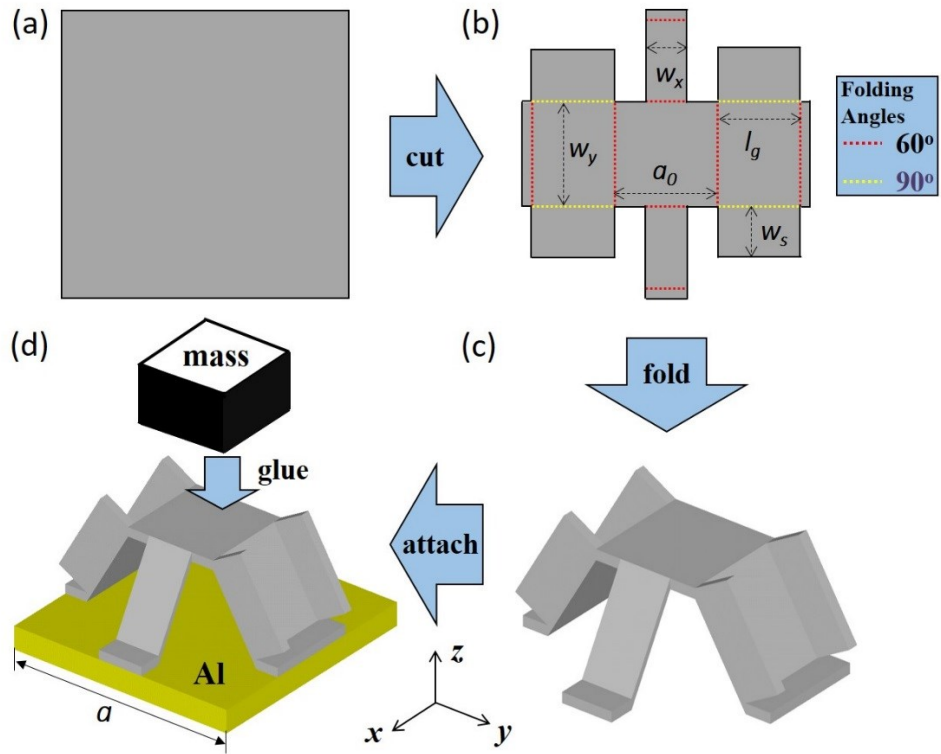

**Figure 1**

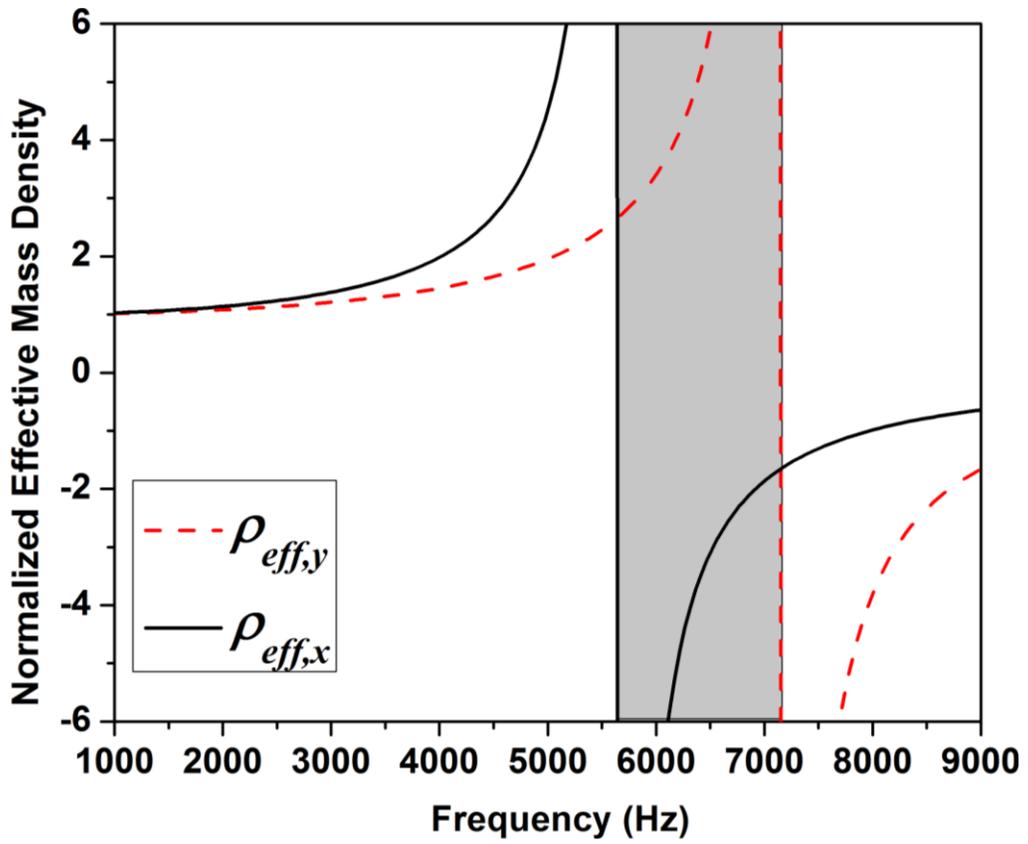
**Figure 2**

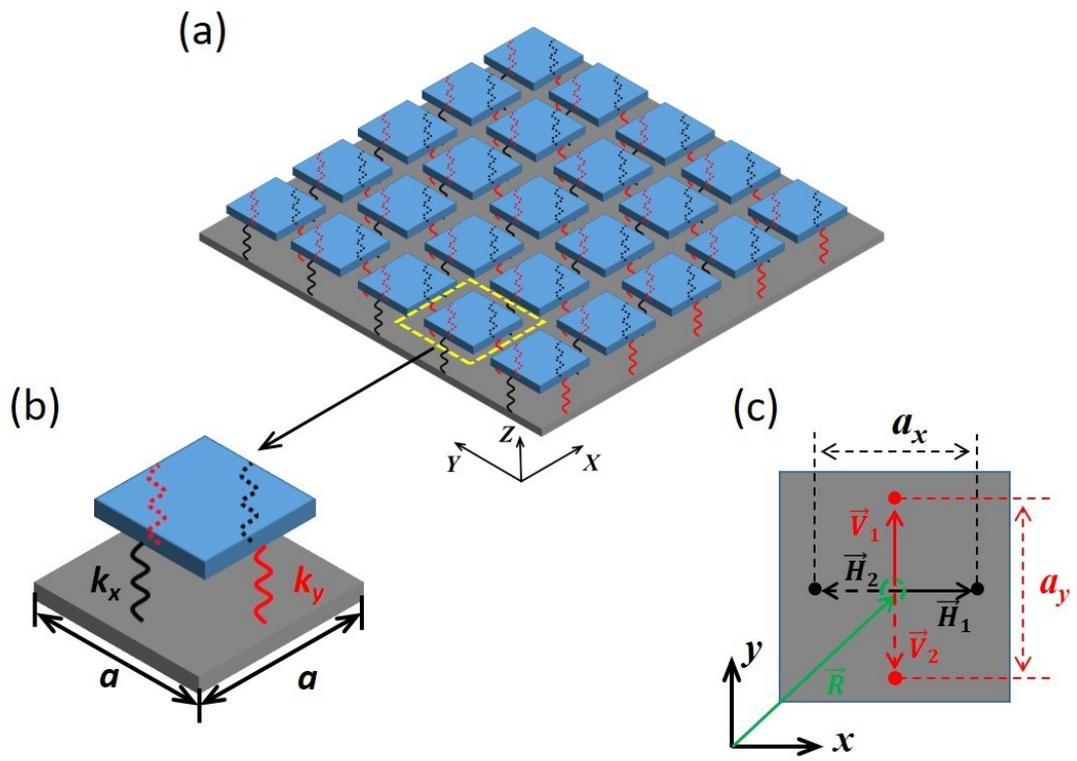

**Figure 3**

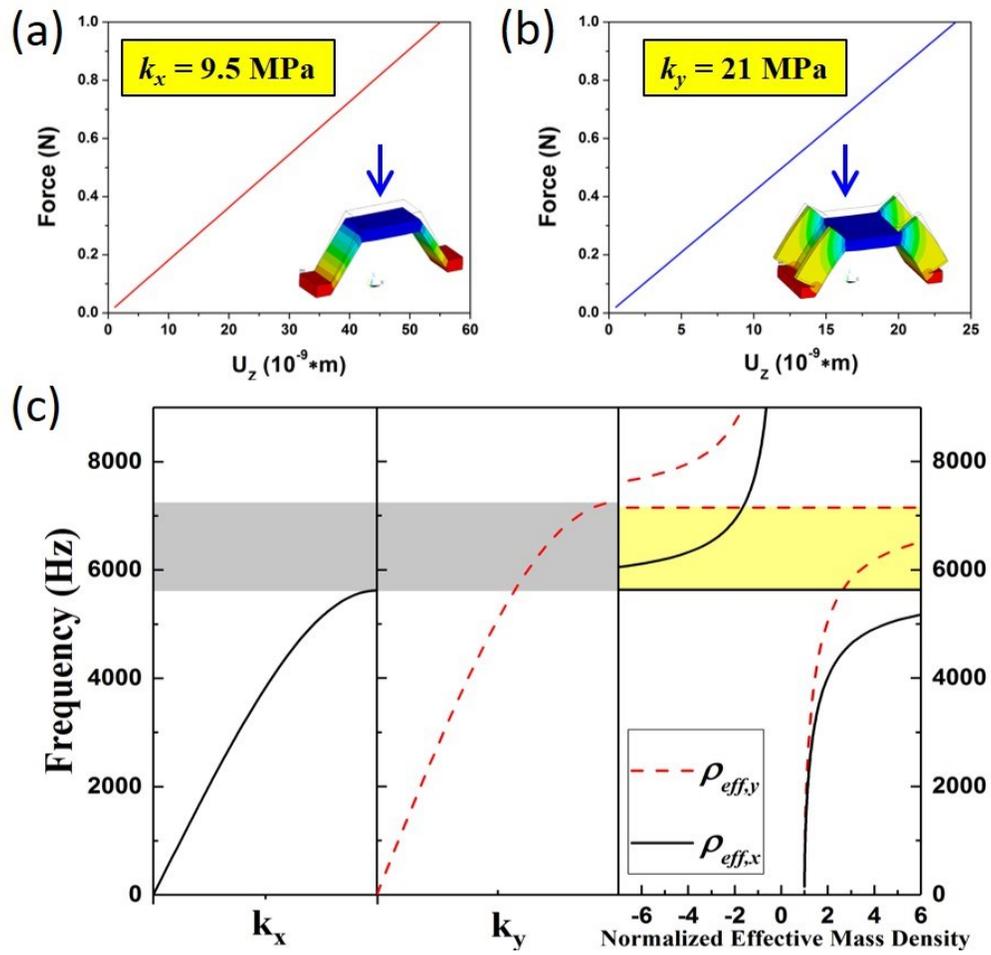

**Figure 4**

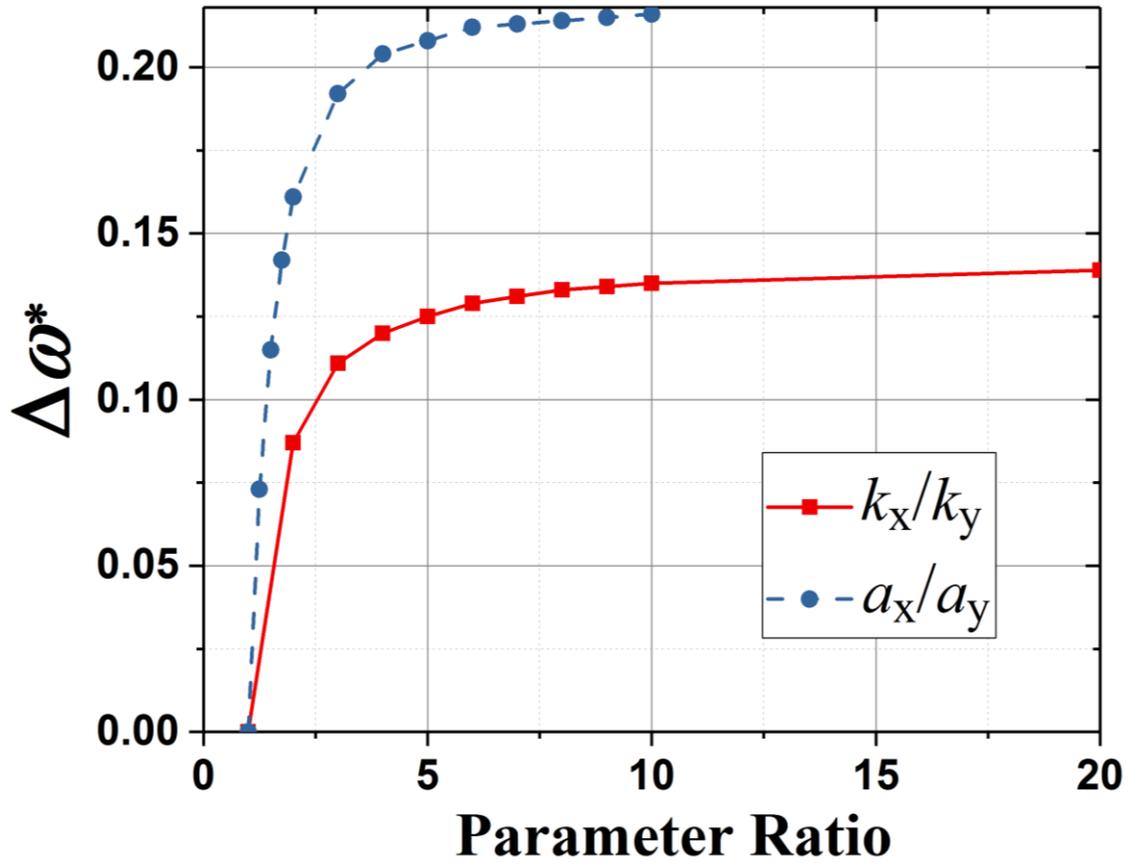

**Figure 5**

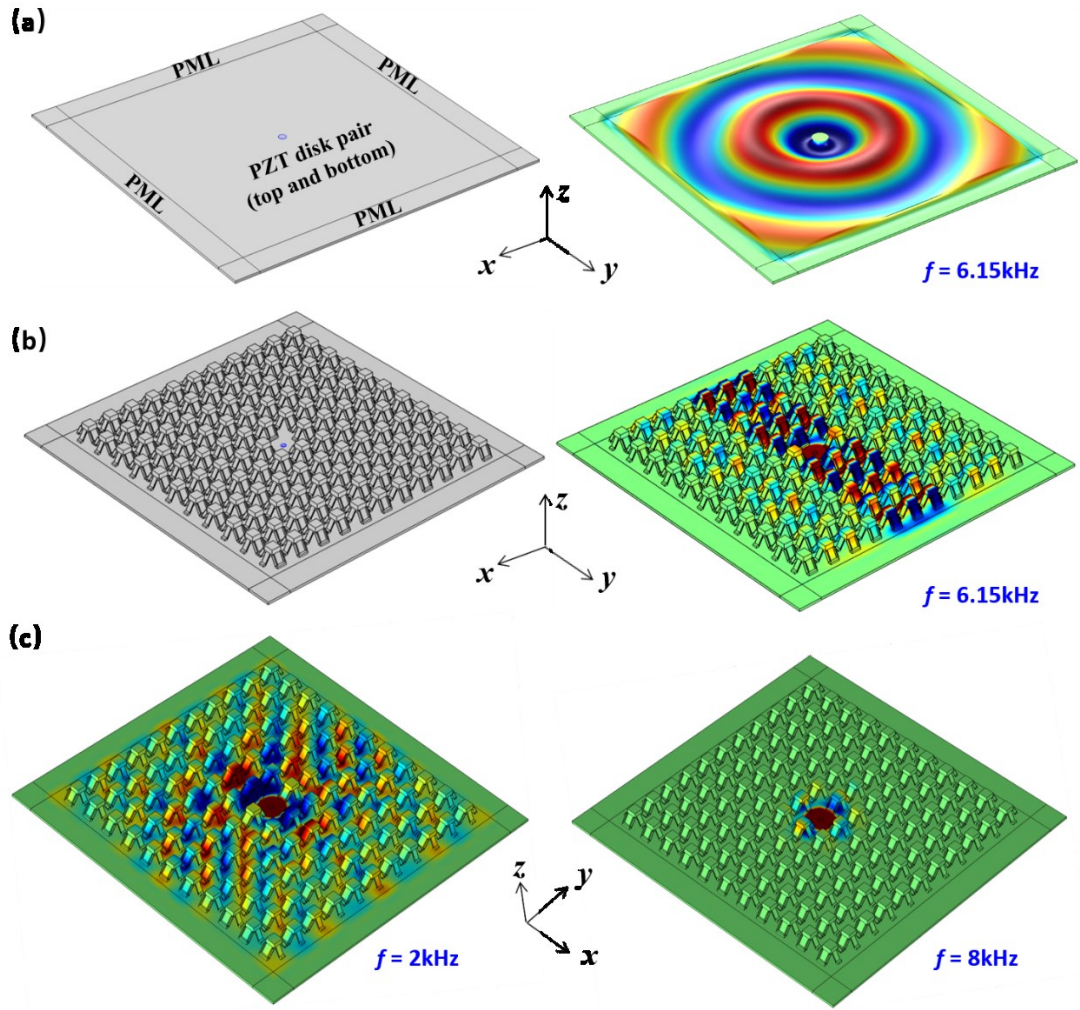

**Figure 6**

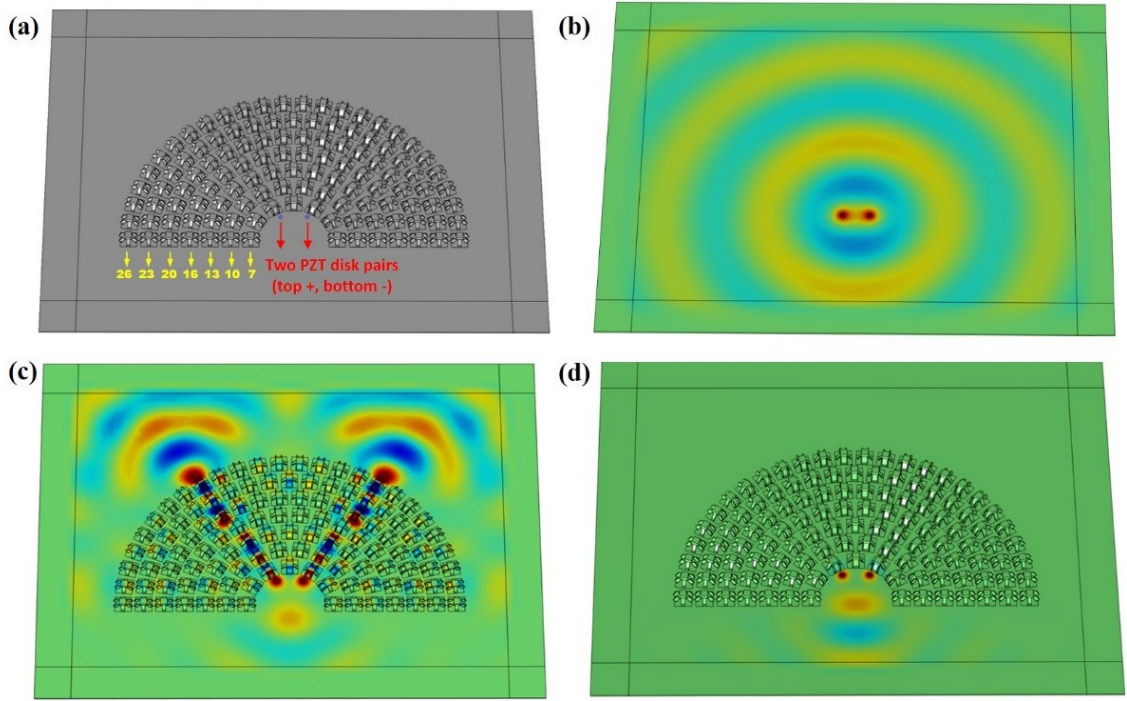

**Figure 7**

# Tables

| Geometrical properties (in mm) | | | | Material properties | |
|---|---|---|---|---|---|
| $a$ | = 10, | $a_0$ | = 4 | Density | =2700kg/m$^3$ |
| $w_x$ | = 2, | $w_s$ | = 2.2 | Young's modulus | = 70 GPa |
| $w_y$ | = 4, | $l_g$ | = 3.6 | Poisson's ratio | = 0.32 |

**Table 1 | The design parameters and material properties of the kirigami structure.**